\documentclass[conference,a4paper,10pt]{IEEEtran}
\usepackage{cite}
\usepackage[T1]{fontenc}
\usepackage[utf8]{inputenc}
\usepackage{csquotes}
\usepackage[english]{babel}
\usepackage{graphicx}
\usepackage{multirow}
\usepackage{tabularx}
\usepackage{booktabs}
\usepackage[caption=false, font=footnotesize]{subfig}
\usepackage{glossaries}
\usepackage{etoolbox}
\usepackage{verbatim}
\usepackage{comment}
\usepackage{microtype}

\usepackage{algorithm2e}

\title{Observations on OMNeT++ Real-Time Behaviour}
\author{%
\IEEEauthorblockN{%
    Christina Obermaier,
    Christian Facchi}
\IEEEauthorblockA{%
    Research Centre, Technische Hochschule Ingolstadt\\
    Email: \{christina.obermaier, christian.facchi\}@thi.de}%
}

\begin{document}
    \maketitle
    \newacronym{etsi}{ETSI}{European Telecommunications Standards Institute}
\newacronym{dcc}{DCC}{Decentralized Congestion Controll}
\newacronym{vanet}{VANET}{Vehicular Ad Hoc Network}
\newacronym{oem}{OEM}{Original Equipment Manufacturer}
\newacronym{hil}{HIL}{Hardware in the Loop}
\newacronym{sil}{SIL}{Software in the Loop}
\newacronym{ivc}{IVC}{Inter-Vehicular-Communication}
\newacronym{c2c}{C2C}{Car to Car}
\newacronym{c2i}{C2I}{Car to Infrastructure}
\newacronym{cam}{CAM}{Cooperative Awareness Message}
\newacronym{ca}{CA}{Cooperative Awareness}
\newacronym{ca}{CA}{Cooperate Awareness}
\newacronym{denm}{DENM}{Decentralized Environmental Notification Message}
\newacronym{btp}{BTP}{Basic Transport Protocol}
\newacronym{ieee}{IEEE}{Institute of Electrical and Electronics Engineers}
\newacronym{its}{ITS}{Intelligent Transport System}
\newacronym{ocb}{OCB}{Outside the Context of a Basic Service Set}
\newacronym{ca}{CA}{Cooperative Awareness}
\newacronym{den}{DEN}{Decentralized Environmental Notification}
\newacronym{c2c-cc}{C2C-CC}{CAR 2 CAR Communication Consortium}
\newacronym{abs}{ABS}{Anti-lock Breaking System}
\newacronym{asr}{ASR}{Anti-slip Regulation}
\newacronym{ecu}{ECU}{Electronic Control Unit}
\newacronym{tc}{TC}{Triggering Condition}
\newacronym{traci}{TraCI}{Traffic Command Interface}
\newacronym{ttc}{TTC}{Time-To-Collision}
\newacronym{irc}{IRC}{Impact Reduction Container}
\newacronym{msdu}{MSDU}{MAC Service Data Unit}
\newacronym{mpdu}{MPDU}{MAC Protocol Data Unit}
\newacronym{mac}{MAC}{Medium Access Control}
\newacronym{txop}{TXOP}{Transmission Opportunity}
\newacronym{ht}{HT}{High Throughput}
\newacronym{dut}{DUT}{Device Under Test}
\newacronym{dll}{DLL}{Data Link Layer}
\newacronym{nic}{NIC}{Network Interface Card}
\newacronym{pdu}{PDU}{Packet Data Unit}
\newacronym{dcc}{DCC}{Distributed Congestion Control}
\newacronym{gps}{GPS}{Global Positioning System}
\newacronym{nmea}{NMEA}{National Marine Electronics Association}
\newacronym{edca}{EDCA}{Enhanced Distributed Channel Access}
\newacronym{csmaca}{CSMA/CA}{Carrier Sense Multiple Access/Collision Avoidance}
\newacronym{dcf}{DCF}{Distributed Coordination Function}
\newacronym{qos}{QoS}{Quality of Service}
\newacronym{ifs}{IFS}{Interframe Space}
\newacronym{difs}{DIFS}{DCF Interframe Space}
\newacronym{pifs}{PIFS}{PCF Interframe Space}
\newacronym{sifs}{SIFS}{Short Interframe Space}
\newacronym{aifs}{AIFS}{Arbitration Interframe Space}
\newacronym{cw}{CW}{Contention Window}
\newacronym{guc}{GUC}{Geographically-Scoped Unicast}
\newacronym{gbc}{GBC}{Geographically-Scoped Broadcast}
\newacronym{gf}{GF}{Greedy Forwarding}
\newacronym{cbf}{CBF}{Contention-Based Forwarding}
\newacronym{c2x}{C2X}{Car to Anything}
\newacronym{c2c}{C2C}{Car to Car}
\newacronym{rsu}{RSU}{Road Side Unit}
\newacronym{ota}{OTA}{Over The Air}
\newacronym{api}{API}{Application Programming Interface}
\newacronym{ada}{ADA}{Advanced Driver Assistance}
\newacronym{ocb}{OCB}{Outside the Context of a Basic Service Set}
\newacronym{traci}{TraCI}{Traffic Control Interface}
\newacronym{ivc}{IVC}{Inter Vehicular Communication}
\newacronym{wave}{WAVE}{Wireless Access in Vehicular Environments}
\newacronym{lpv}{LPV}{Long Position Vector}
\newacronym{ack}{ACK}{Acknowledgement Frame}

    \begin{abstract}
    \emph{OMNeT++} is a widely used platform for all types of network simulations.
    The open source simulation framework \emph{Artery} can be used to perform \gls{vanet} simulations.
    This paper presents an approach for connecting this simulation and real-world \gls{vanet} hardware to extend the test range and investigates the real-time behaviour of the simulation.
    As a \gls{dut} depends on real-time data to perform properly, different simulation scenarios running different hardware setups are presented.
    Additionally, the paper deals with the impacts of real-time losses on the test run outcomes.
    Most time dependant algorithms like the duplicate packet detection do not need very accurate real-time data and thus could be verified using the presented approach.
    Otherwise, in some cases such as testing of multi-hop communication, accurate real time is crucial.
    \end{abstract}
    \section{Introduction}
In times of increasing complexity of advanced driver assistance systems, it is crucial to enhance the environmental awareness of vehicles.
Vehicles can be equipped with \glsentryfull{vanet} devices, acting as a new information source besides already well known sensors.

\Glspl{vanet} are spontaneously created networks between road participants and \glspl{rsu}.
They are based on IEEE 802.11p \cite{IEEE80211} and ETSI ITS G5 \cite{its10} in Europe as well as IEEE WAVE in the USA \cite{wave14}.
This paper focuses on the European ETSI ITS G5 standards.

\Glspl{vanet} and their applications were developed with the focus on enhancing traffic safety and traffic flow \cite{Wil09}.
Especially in critical driving situations, availability of information is crucial.
This leads to the question, how to test \gls{vanet} communication properly.
As \gls{sil} simulation is not enough to ensure the availability, this paper evaluates if \emph{Artery} can fulfil the real-time requirements necessary for \gls{hil} testing.

In section \ref{sec:relatedWork} a overview of the related testing frameworks and hardware testbeds is given.
Section \ref{sec:concept} presents the state of the art and the theoretic concept of the hardware testbed.
In section \ref{sec:realTime} different scenarios running on different hardware setups are investigated.
Section \ref{sec:evaluation} presents the limitations of the presented approach.
Finally, section \ref{sec:conclusion} includes a conclusion and a brief outlook.

\section{Related Work}
\label{sec:relatedWork}
Currently there are mainly two different \gls{ivc} testing approaches.
On the one hand, there is pure software testing with frameworks like \emph{Artery}\footnote{https://github.com/riebl/artery} \cite{Rie15} and \emph{Veins}\footnote{http://veins.car2x.org} \cite{Som11} which are based on the discrete event simulator \emph{OMNeT++}.
These implement the \gls{its} G5 and the \gls{wave} standards, respectively.
Also, different environmental circumstances and accidents can be modelled easily \cite{Obe17}.

On the other hand, there are real-world testing approaches like the "Testfeld A9" established near Ingolstadt in Germany.
Real-world testing allows for testing actual \gls{ivc} hardware, but it is required to have at least two drivers in real vehicles equipped with \gls{ivc} hardware.
Thus, field tests are hard to reproduce and very expensive.
While it is still possible to perform simple real-world scenarios, like presented in \cite{Lee07}, it might be not feasible to do this with more complicated scenarios.

\Gls{hil} tests can be used to perform hardware tests which are easy to repeat and cost effective.
It is not a new approach using \emph{OMNeT++} to connect a simulation with real-world hardware as \emph{OMNeT++} was one of the fastest simulators in the domain of wireless networks in prior software versions \cite{Khan12}.
In \cite{Siv14}, a routing framework for \emph{OMNeT++} \gls{hil} simulations is presented and the real-time behaviour of \emph{OMNeT++} was investigated in different scenarios.
They mentioned that real time will be a problem in future scenarios, especially if the node topology begins to change dynamically due to node mobility.
According to \cite{Boehm15}, \emph{OMNeT++} can also be coupled with \emph{RoSeNet} to connect with real-hardware sensors.
In \cite{Aus15} the performance of the \emph{INET} framework (release 2.5.1) in emulation mode combined with an enhanced real-time scheduler was investigated.
They pointed out to have performance problems.
Additionally, they observed packet losses occurring in the communication between the real hardware and the simulation.
\section{HiL concept}
\label{sec:concept}
\emph{Artery} is an open source framework for the discrete event simulation \emph{OMNeT++} \cite{Rie15}.
It allows for simulating European \glspl{vanet} using \emph{Vanetza}\footnote{http://vanetza.org/}, which is an open source implementation of the ETSI \gls{its} G5 communication stack \cite{Rie17}.
\emph{Veins} or \emph{INET} provide the physical and \gls{mac} layers.
Moreover, the movement of the vehicles is simulated by the open source traffic simulator \emph{SUMO}\footnote{http://sumo.dlr.de/}.
\emph{SUMO} and \emph{Artery} are coupled using the \gls{traci} \cite{Wegener2008}.

\subsection{Artery Overview}
Basically, each vehicle controlled by \emph{SUMO} is represented by an \emph{OMNeT++} compound-module called \emph{Car}.
A \emph{Car} module is, among others, composed of a \emph{Middleware}, a \emph{VanetNic} and a \emph{Mobility} submodule.
The \emph{Mobility} module is responsible for all vehicle dynamics related data and information.
The \emph{VanetNic} represents the network interface of each car.
The \emph{Middleware} module hosts all registered applications and contains an instance of \emph{Vanetza}, taking care of routing and transport the incoming and outgoing \gls{ivc} packages.
Currently, the day one applications \gls{cam} and \gls{denm} are provided by \emph{Artery} \cite{Rie15, Artery}.

\Gls{ca} is responsible to inform all road users in the network about basic data of all other participating vehicles.
This basic data contains, among others, the network node's position, speed and heading.
\Glspl{cam} are generated up to ten times per second from each road participant \cite{its14Cam}.
In contrast, \glspl{denm} are only generated in case of a specific event.
A \gls{denm} contains information of hazards and dangerous situtations occurring in road transport \cite{its14Denm}.
Thus, \emph{Artery} allows for defining this dangerous situations and environmental influences like, for example, heavy rain or nearly accidents \cite{Obe17}.
These situations are required to trigger related \gls{denm} messages.

\subsection{HiL concept}
\label{subsec:hilConcept}
\begin{figure}
    \centering
    \includegraphics[width=\columnwidth]{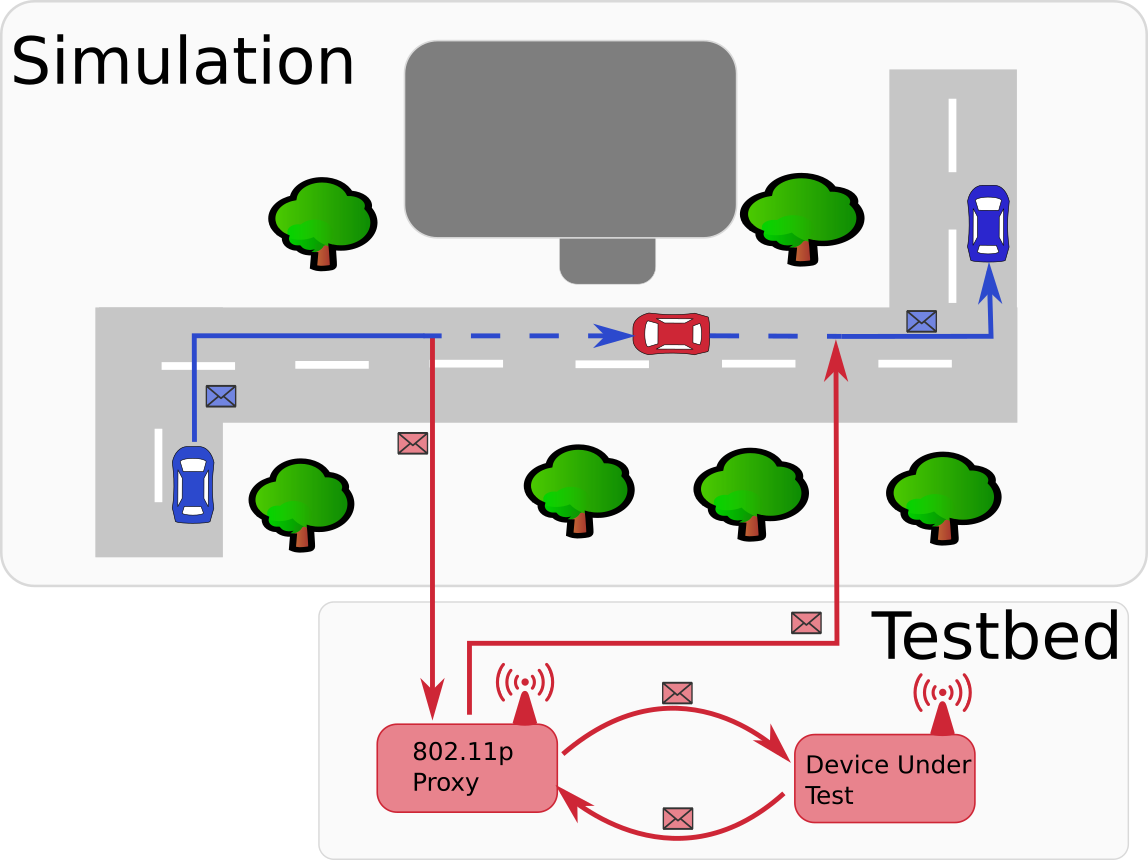}
    \caption{HiL concept overview}
    \label{fig:overview}
\end{figure}
Figure \ref{fig:overview} shows the basic concept of a \gls{hil} simulation using \emph{Artery} to provide test data.
The red coloured vehicle is the simulated representative of a \glsentryfull{dut}.
For now, this car will further be called physical twin.
If a message is going to be transmitted to the physical twin, the message will be transferred to a 802.11p gateway realised on software defined radio technology.
This behaviour is represented by red arrows, showing the connection of the simulation to the real word and thus with the \gls{dut}.
The blue dotted arrows show the current behaviour of the simulation.

In contrast to other simulated cars, the physical twin has a quite limited functionality:
Upper protocol layer processing is done by the \gls{dut} so the \emph{Middleware} module as well as the \emph{Vanetza} instance can be dropped.
Otherwise, the \emph{Mobility} module is still needed to provide \gls{gps} data to the \gls{dut} to ensure properly working routing protocols \cite{its14}.
Without appropriate \gls{gps} information, the geographic routing protocols defined in \cite{its14} would not work properly.
Summarising, all functionality beginning with the \gls{mac} layer processing is stripped from the physical twin.
It is only responsible for calculating message receptions and feeding received messages back on the \emph{OMNeT++} channel.

As each kind of hardware testbed is dependent on real-time execution, the simulated environment must run in real time, too.
To ensure this, the \emph{OMNeT++} built-in scheduler is exchanged by a real-time scheduler.
This scheduler is built upon \emph{Boost ASIO} timers, ensuring asynchronous waiting.
Hence, the real-time scheduler slows down the simulation, if it could run faster than real time.
Additionally, it is aware of real-time losses so that it could stop the simulation if data could not be provided in real time.
Also, the scheduler provides logging mechanisms to investigate the real-time behaviour of simulation runs after they are finished.
A pseudocode implementation of the scheduler can be found in Appendix \ref{appendix}.

\section{Investigations on Real-Time Behaviour}
\label{sec:realTime}
\begin{table}[t]
	\centering
    \caption{Hardware setup}
        \begin{tabularx}{\columnwidth}{c | X | X }
    	Component & Laptop Computer & Simulation Cluster \\
        \toprule
        CPU & Intel Core i5-6300U @ 2.40GHz & Intel Xeon E7-8867 v4 @ 2.40GHz\\
        \midrule
        Cores & 1 x 4 & 4 x 18 \\
        \midrule
        RAM & 16GB & 3TB \\
        \midrule
        Hard Drive & 256GB SSD & 450GB SAS SSD RAID 1\\
	\end{tabularx}
    \label{tab:hardware}
\end{table}
\begin{table}[t]
	\centering
    \caption{Event Mapping}
        \begin{tabularx}{\columnwidth}{c| p{90px} | X | X}
    	ID & Event name & \# Events \newline "3 vehicles" & \# Events \newline"5 vehicles" \\
        \toprule
        1 & TraCI Connect & 1 & 1\\
        \midrule
        2 & TraCI Step & 322 & 370\\
        \midrule
        3 & GeoNet packet & 3870 & 11298\\
        \midrule
        4 & GeoNet data frame & 3870 & 11298\\
        \midrule
        5 & txStart-0 & 3 & 5\\
        \midrule
        6 & endIFS & 661 & 1189\\
        \midrule
        7 & configureRadioMode & 1322 & 2378\\
        \midrule
        8 & transmissionTimer & 661 & 1189\\
        \midrule
        9 & remove non Interfering Transmission & 661 & 1188\\
        \midrule
        10 & report CL & 928 & 1650\\
        \midrule
        11 & middleware update & 925 & 1645\\
        \midrule
        12 & txStart-1 & 658 & 1184\\
        \midrule
        13 & GeoNet radio frame & 1274 & 4460\\
        \midrule
        14 & reception Timer & 1274 & 4460\\
        \bottomrule
        & Overall events & 16430 & 42315 \\
	\end{tabularx}
    \label{tab:events}
\end{table}

As already known from investigations described in \cite{Neu15} there is a quite low execution speed of \emph{OMNeT++} \gls{vanet} simulations.
This leads to the question, if \emph{Artery} is capable of reaching and holding real time to provide data for a \gls{hil} testbed while doing an online simulation.
\subsection{Scenario Description}
The chosen scenario is very simple:
Three vehicles driving on a highway from north to south.

As the amount of driving vehicles seems to be the main influence on the execution speed of these simulations, the second test scenario increases the amount of driving vehicles by two.
All other parameters remain the same as in the previous scenario.
The setup of the radio medium is configured by \emph{Artery} using its \emph{INET} defaults.
The in Section \ref{subsec:hilConcept} introduced scheduler is used to evaluate the real-time behaviour.
The simulation was built using \emph{OMNeT++} version 5.1.1 and was executed on two different computers: A laptop computer and a simulation cluster.
Table \ref{tab:hardware} presents a comrehensive hardware list of the used computers.
As \emph{OMNeT++} uses only one core in this test setup, the speed-up of the simulation running on the simulation cluster is caused by faster calculation hardware not by better parallelism.

\subsection{Simulation Results and Scenario Comparison}
\begin{figure*}
\begin{minipage}[t]{1.0\columnwidth}
    \subfloat[\label{fig:asyncBaseIconic_boxplot} Average event times]{
        \includegraphics[width=0.99\columnwidth]{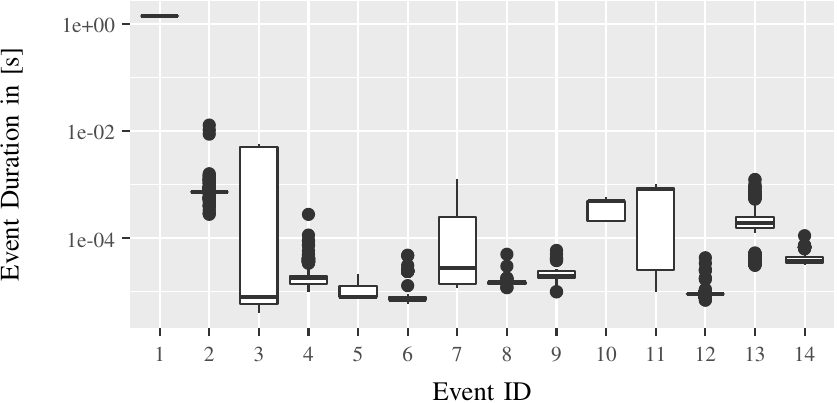}
    }
    \quad
    \subfloat[\label{fig:iconic-base_histogram} Amount of critical real-time losses in base scenario]{
        \includegraphics[width=0.99\columnwidth]{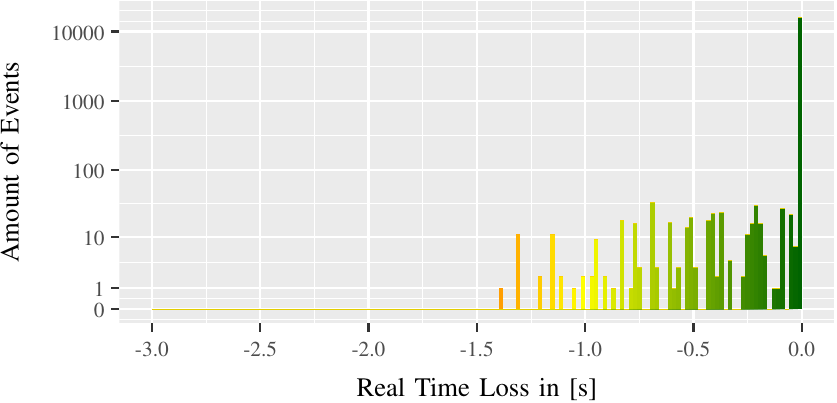}
    }
    \quad
    \subfloat[\label{fig:iconic-five_histogram} Amount of critical real-time losses in five cars scenario]{
        \includegraphics[width=0.99\columnwidth]{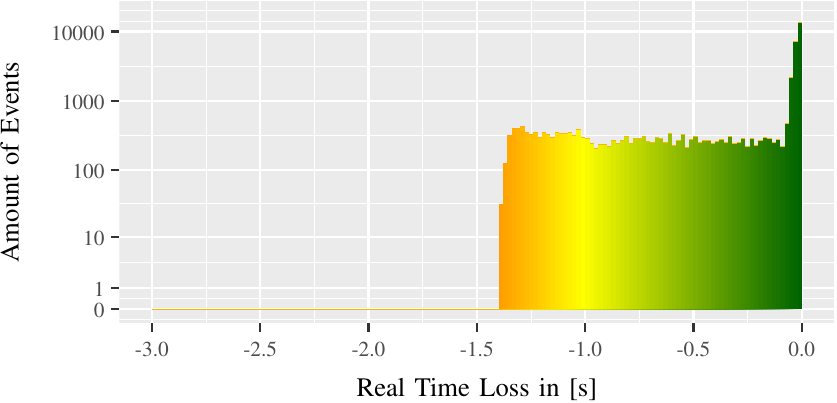}
    }
    \caption{Scenarios executed on the simulation cluster}
    \label{fig:iconic}
\end{minipage}
\begin{minipage}[t]{1.0\columnwidth}
    \subfloat[\label{fig:asyncMoreCars_boxplot} Average event times]{
        \includegraphics[width=0.99\columnwidth]{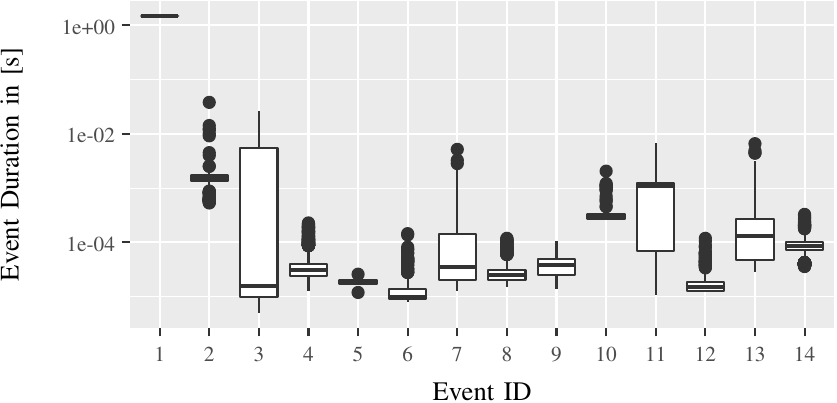}
    }
    \quad
    \subfloat[\label{fig:laptop-base_histogram} Amount of critical real-time losses in base scenario]{
        \includegraphics[width=0.99\columnwidth]{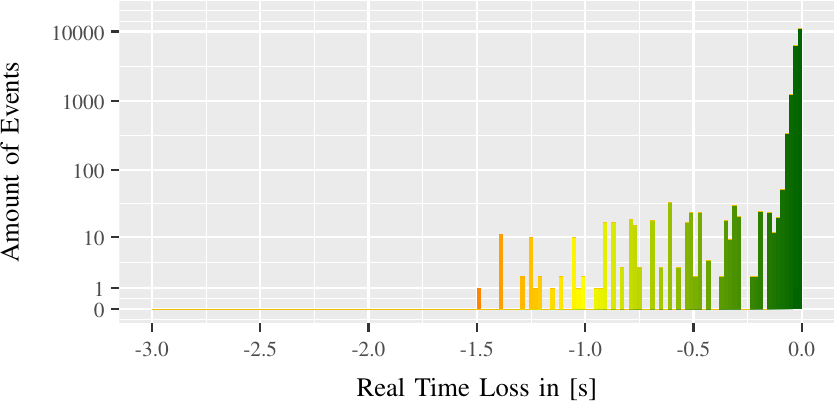}
    }
    \quad
    \subfloat[\label{fig:laptop-five_histogram} Amount of critical real-time losses in five cars scenario]{
        \includegraphics[width=0.99\columnwidth]{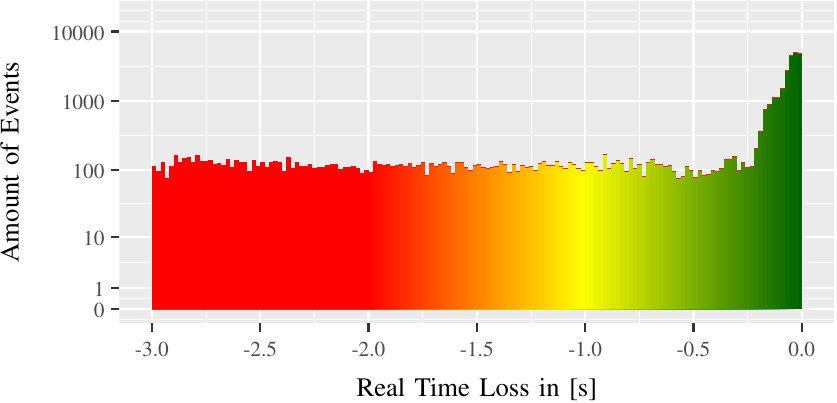}
    }
    \caption{Scenario executed on the laptop computer}
    \label{fig:laptop}
\end{minipage}
\end{figure*}

\begin{figure*}
	\centering
    \subfloat {
    \includegraphics[width=0.985\textwidth]{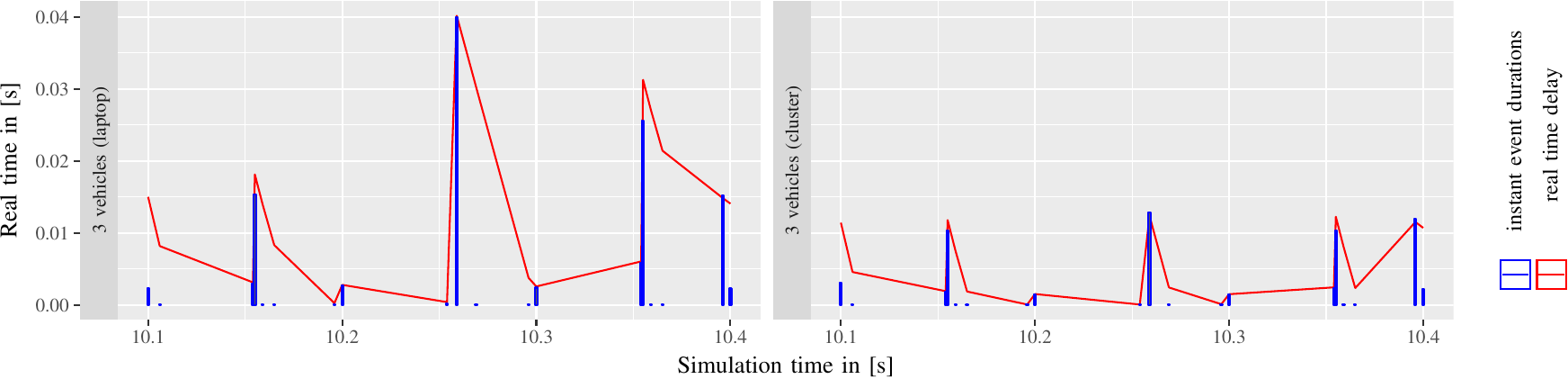} }
    \quad
    \subfloat {
    \includegraphics[width=0.985\textwidth]{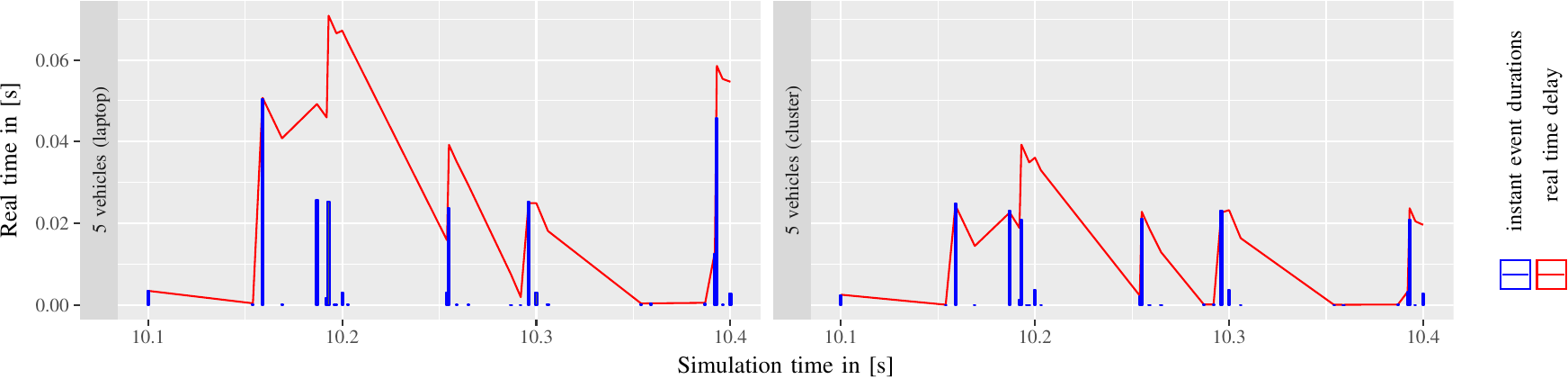} }

    \caption{Real-time delays and event durations}
    \label{fig:all_lineChart}
\end{figure*}

Table \ref{tab:events} presents the number of events occurring in both scenarios.
The ID column is used to map event IDs to the events depicted in the box plots in figure \ref{fig:asyncBaseIconic_boxplot} and \ref{fig:asyncMoreCars_boxplot}.
The third and the fourth columns show the number of occurring events dependent on the simulated scenario.
Thus, compared to the base scenario which triggers 16430 events, the scenario with two more vehicles triggers 42315 events.
Hence, adding two cars causes 2.58 times more events in this scenario setup.

Figure \ref{fig:iconic} and \ref{fig:laptop} present the evaluations of both scenarios executed on the simulation cluster and the laptop computer.
The included boxplots depict the execution time of events.
The indices on the x-axis correspond to the events mentioned in Table \ref{tab:events}.
If we compare both figures, the time a event needs to be executed is lower while using the simulation cluster but relative event durations remain nearly the same.
This behaviour is caused by the faster computing capabilities of the single cores of the simulation cluster.

The histograms \ref{fig:laptop-base_histogram} and \ref{fig:laptop-five_histogram} show the real-time misses per event while the scenarios were executed on the laptop.
Histogram \ref{fig:iconic-base_histogram} and \ref{fig:iconic-five_histogram} present the same but for the execution on the simulation cluster.
As the green bars indicate events executed nearly in real time, higher green bars and lower red bars indicate a nearly real-time capable simulation run.

It can be seen that there are real-time drops up to 1.5 seconds in each scenario.
This is caused by the \emph{TraCi Connect} event, which is executed as first event and takes about one second.
Thus, even if the scenario itself is real-time capable, there will be always a few seconds at the beginning of the simulation which must be skipped because of the system startup.
Figure \ref{fig:iconic-base_histogram} is a great example for a good and fast running scenario.
There are only roughly 50 - 100 events which are more than one second behind the real time.
About 15000 events out of the total amount of 16430 events are executed nearly in real time.
This means, the simulation cluster can basically handle a small scenario with three vehicles in real time.

Figure \ref{fig:laptop-base_histogram} presents the same scenario executed on a laptop computer.
It can be seen that there are still about 10000 events executed nearly in real time.
Other 5000 events are executed only 0.1 second behind real time.
Thus, only a few more events can be found in the area of about one second behind real time, but the scenario looks still quite good.

Distinct differences between execution speed of the computing hardware can be seen in Figure \ref{fig:iconic-five_histogram} and Figure \ref{fig:laptop-five_histogram}.
The simulation cluster is able to handle the five car scenario with still up to 1.5 seconds real-time loss.
The amount of events in the area from 1 to 1.5 seconds is significantly increased, compared to the base scenario but \emph{OMNeT++} is still able to catch up.
However, the five car scenario executed on the laptop computer exceeds the 1.5 seconds limit by far.
There are many real-time losses up to three seconds.
Thus, the five car scenario can only be handled by the simulation cluster.

Figure \ref{fig:all_lineChart} admits a closer look on the real-time behaviour of the different simulation runs.
Blue bars show the time needed to process all events occurring at a particular simulation time stamp.
The red line depict the current gap between simulation time and real time.
Thus, a higher blue bar causes a higher real-time loss peak.
It seems that \emph{Artery} tends to schedule a bunch of events every 50ms.
This behaviour causes the simulation to fall behind real time every 50ms, which is indicated by the peaks of the red line.

As it can be seen, the simulation cluster produces as many real-time losses as the laptop computer.
But, even if the differences between the execution time of single events are not that clear, the real-time losses are much higher on the laptop computer, independent from the executed scenario.
This approves that \emph{Artery} has to execute many events at the same time because they are scheduled at nearly the same timestamp.
Due to the lack of parallelism, \emph{OMNeT++} has to execute this events in a sequence, which causes the real-time losses.

As there are many events which are closely linked together, for example the \emph{GeoNet packet} and the \emph{GeoNet data frame}, this behaviour could not be changed significantly.
Thus, there will always be real-time losses caused by many events occurring at the same time if there is no parallelism.

\subsection{OMNeT++ Time Flow}
\begin{figure}[t]
    \centering
    \includegraphics[width=0.9\columnwidth]{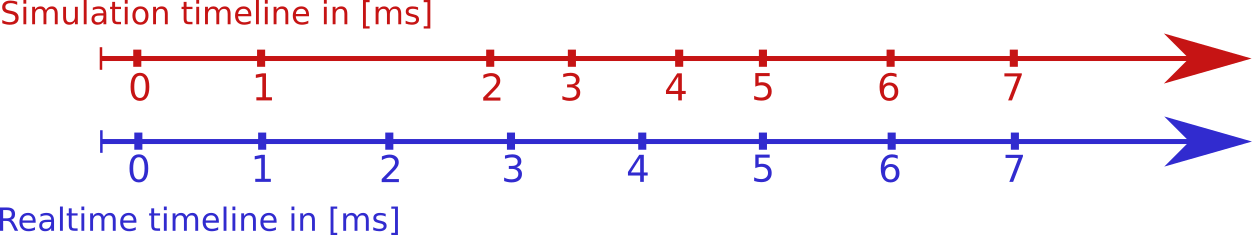}
    \caption{OMNeT++ and real-time timeline}
    \label{fig:timeline}
\end{figure}

Figure \ref{fig:timeline} depicts a generic example how the timeline in \emph{OMNeT++} behaves compared to the real-time flow.

As it can be seen, the absolute time needed to execute one simulation millisecond is sometimes higher than one real millisecond like indicated at millisecond two.
If the simulation is behind the wall-clock time, it has to catch up to avoid higher real-time gaps.
Hence, the simulation is trying to execute events faster than in real-time till simulation time and wall-clock time are matching again.
But, even if the \emph{OMNeT++} clock sometimes progresses faster than real time, the fixed time stamp is always behind or equal to the wall-clock time which is ensured by the real-time scheduler.
If the next event is located in the future, the scheduler waits till the timestamp of the event matches the wall-clock time.
Thus, it is ensured that the \gls{dut} must only deal with timestamps in the past and not in the future, which would be much more problematic.

\section{Simulation Result Interpretation}
\label{sec:evaluation}
Most scenarios in \gls{ivc} communication do not depend on very accurate timestamps.
For example, the \gls{lpv} \cite[Section 8.8]{its14} contains the time at which the last geographic position update was received by the vehicle writing the \gls{lpv}.
As in a real-world environment \gls{gps} update rates are fluctuating, the timestamp in the \gls{lpv} is not strongly related to the current wall-clock time.

The \gls{lpv} timestamp is used to perform duplicate packet detection at the network layer \cite[p. 63]{its14}.
The duplicate packet detection algorithms depend on the sequence number as well as the timestamp.
A packet is identified to be a duplicate if the timestamp is lower or equal to an already received packet.
As the introduced \emph{RealTimeScheduler} ensures that the simulation time is always behind or equal to the wall-clock time, this algorithm works like expected.

\Glspl{cam} contain a generation timestamp \cite{its14Cam} used, for example, to recognise and avoid replay attacks \cite[Section 7.6.1]{its10}.
However, a \gls{cam} is not detected to be a duplicate if there is only a slight difference between wall clock time and the generation timestamp.

A \gls{denm} contains various timestamps as well.
There is, for example, an expiry time after which a \gls{denm} event is terminated.
Mostly, this period lasts several seconds but in cases like the dangerous situation \cite{TrigDanger} trigger, a \gls{denm} event lasts only two seconds.
Thus, in some situations a real-time loss higher than one ore two seconds may invalidate a \gls{denm} wrongly.

Another issue is the long-range communication using multiple vehicles as forwarders \cite{its14}.
A \gls{mac} layer unicast is followed by an \gls{ack}, sent by the receiver, to confirm a successful packet transmission.
The time span in which the \gls{ack} is expected by the initial sender is the time of a \gls{sifs} \cite[Section 9.3.2.8]{IEEE80211}.
Determined by the channel bandwith of 10 MHz \cite{its13} the \gls{sifs} is 32 $\mu s$ \cite[Table 18-17]{IEEE80211}.
This very short time span leads to a potential problem:
If a MAC layer unicast is sent to the \gls{dut}, it has to respond with an \gls{ack} within this \gls{sifs}.
As depicted in Figure \ref{fig:timeline} between millisecond two and three, \emph{OMNeT++} could run faster than real time.
Thus, the sending vehicle inside \emph{OMNeT++} may not receive this message in the claimed timespan.
Hence, this problem must be solved to perform multi-hop tests.

Summarizing, in most cases, a real-time loss in a range of a few milliseconds is not that problematic.
Important algorithms like the duplicate packet detection do only rely on linear time flow and do not depend on hard real time.
Only a few scenarios like multi-hop testing relay on a very accurate time synchronisation.
Thus, \emph{Artery} is basically able to provide online simulation data for a real-time testbed.

\section{Conclusion}
\label{sec:conclusion}
This paper presented an approach for extending \emph{Artery} to provide hardware tests.
\emph{Artery} is used to facilitate \gls{sil} tests in the area of \glspl{vanet}.
As a \gls{hil} simulation always depends on real-time data, the online simulated scenarios have to be executed in this manner.
It was investigated that this criteria can only be fulfilled if only a few cars are simulated.
Also, even if the simulation is overall real-time capable, there are always real-time losses.
How significant these losses are depends on the used computation hardware and the scenario complexity.
As \emph{OMNeT++} only uses one core when simulating \glspl{vanet}, a higher single core performance causes a higher execution speed.
Also, \emph{Artery} and other wireless communication models tends to produce events to be executed nearly at the same time, causing more significant real-time losses.
This is related the fact that one sending event triggers various receiving events.

Also, it was audited in which situations nearly hard real time is required and when real-time drops are bearable:
Multi-hop communication strongly depends on real time data, so its not possible to test this feature properly in the current state of the simulation.
Most other algorithms do only depend on a steady time flow and are not influenced by real-time losses in the range of a few milliseconds.
This is the case for \gls{denm} message expiries or the recognition of replay attacks.
Also the duplicate packet detection is not affected harmfully.

In conclusion, \emph{Artery} could basically be used to provide data for \gls{vanet} \gls{hil} tests.
In case of multi-hop-test scenarios, real-time losses may influence the test results heavily.
Other scenarios can be used to provide functional testing of \gls{vanet} hardware.
This scenarios are currently limited to three or four vehicles simulated vehicles depending on the used hardware.

As this paper presents a work in progress research project, the \gls{hil} testbed will be created with the observed behaviour of \emph{OMNeT++} in mind.
Thus, to enable multi-hop testing, which is crucial in European \glspl{vanet}, one idea is to use a Software Defined Radio (SDR) as 802.11p proxy.
This allows for a modified \gls{mac} layer of the used proxy device to send ACK frames depending on the current state of the simulation.
Hence, the proxy device must know all vehicles which can communicate with the \gls{dut} in the current simulation step.
This idea could provide a basic implementation of a testbed which can handle most functionality of ETSI \gls{its} G5 networks with the constraint that the simulation must run in nearly real time.

So, further work must be done in the field of enhancing simulation speed to achieve faster running simulation scenarios.
Moreover, other ways to provide test data for hardware tests can be investigated.
This includes, among others, the capturing of the simulated network traffic and playing them back in real time.

\appendices
\section{Real Time Scheduler}
\label{appendix}
\begin{algorithm}
\SetAlgoLined
\KwResult{next cEvent}
 \emph{currentRealTimeMiss} = \emph{simTime} - \emph{wallClockTime}\;
 \eIf{(currentRealTimeMiss * -1) > realTimeMissThreshold}{
  \tcp{simulation unacceptable slow}
  stop simulation\;
 }{
 \emph{eventDuration} = \emph{wallClockTime} - \emph{eventStartTime}\;
 log \emph{currentRealTimeMiss} and \emph{eventDuration} and \emph{nextEventIdentifier}\;
 \While{SimTime > wallClockTime}{
  \tcp{simulation faster than real time}
  wait\;
 }
 set \emph{nextEventIdentifier}\;
 set \emph{eventStartTime}\;
 return \emph{nextEvent}\;
 }
 \caption{\emph{cEvent* RealTimeScheduler::takeNextEvent} pseudocode}
\end{algorithm}

	\bibliographystyle{IEEEtran}

\end{document}